# Data-Plane Security Applications in Adversarial Settings

Liang Wang
Princeton University
lw19@princeton.edu

Prateek Mittal
Princeton University
pmittal@princeton.edu

Jennifer Rexford
Princeton University
jrex@cs.princeton.edu

## ABSTRACT

High-speed programmable switches have emerged as a promising building block for developing performant data-plane applications. In this paper, we argue that the resource constraints and programming model in hardware switches has led to developers adopting problematic design patterns, whose security implications are not widely understood. We bridge the gap by identifying the major challenges and common design pitfalls in switch-based applications in adversarial settings. Examining six recently-proposed switch-based security applications, we find that adversaries can exploit these design pitfalls to completely bypass the protection these applications were designed to provide, or disrupt system operations by introducing collateral damage.

## 1 INTRODUCTION

Programmable data-plane hardware and programming languages such as P4 [1, 26] are enablers for performant in-network applications. Recently proposed applications have leveraged high-speed programmable switches to support a diverse set of security tasks (spoofing detection, covert-channel detection, DDoS prevention, intrusion detection, etc.) [2, 19, 23, 30, 38, 39]. Compared to end-host-based security solutions, data-plane security applications have the potential to protect all end-hosts in a network *at high speed*, without modifications to the end-hosts.

However, due to the limited memory and processing resources, and the constrained programming model, programmable data planes introduce a different design space than traditional software systems. For instance, to achieve line-rate performance in data planes, developers commonly choose to implement core functionality only without considering corner cases (i.e., *reduction* operations) or replace a secure hash function with a weak hash (i.e., *approximation* operations). Nevertheless, such reduction and approximation may compromise the very security properties a security application is trying to assure in adversarial settings. While the networking and security communities are actively exploring security use cases of programmable data planes, surprisingly little attention has been paid to the security implications of the new design patterns.

In this work, we discover security failures in six programmable-hardware-switch-based (or switch-based for simplicity) security applications (NetHCF [38], NetWarden [38], two SYN proxies [23, 30], P4RTT [2], and P4 Knocking [39]) and analyze the potential causes. We find that the fundamental limitations in (1) in-network monitoring (e.g., no efficient spoofing detection mechanisms and no visibility into end-host state) and (2) resource constraints in switch hardware (necessitating performance optimizations in switch-based applications) have led to the adoption of several problematic design patterns. Exploiting the identified design pitfalls, we find all of the examined applications are vulnerable to our novel evasion (i.e., evading the detection of the application) or disruption (i.e., exploiting the system to disrupt other users' traffic) attacks under their original threat models. Our key results are as follows:

- Exploiting the limited visibility of end-host state in switch-based applications, the adversary can trick NetHCF (an in-network spoofing-detection system) into believing that legitimate communications are happening between a spoofed IP and a protected host, which is a stepping stone for launching evasion and disruption attacks.
- NetWarden, a covert-channel detection system, can be bypassed by manipulating payload-related packet headers (e.g., checksum), which current programmable switches cannot easily validate.
- The adversary can bypass known switch-based SYN proxies and launch SYN flooding attacks with minimal effort, due to their use of CRC as a hash function for generating cookies.

Additionally, we demonstrate that certain design choices may cause deployability issues (e.g., NetWarden can disrupt 18% of HTTP(S) flows), and may also enable side-channel attacks that can reveal sensitive information.

One fundamental motivation for adopting those problematic design patterns is optimizing application performance (e.g., achieving line-rate performance) on switch data planes. We argue that such a performance-prioritized approach to application design is severely flawed for switch-based security applications. Instead, security should be prioritized over performance, considering the adversarial context in which security applications operate. A related problematic trend is for application designers to present an *incomplete* design, which ignores certain security issues or assumes they can be fixed with certain "patches" to the applications [17, 19, 38, 40]. Not only does this undermine security, an incomplete design will not allow the community to fairly evaluate application performance, as the expected performance results may significantly change after patching.

Towards building practical and deployable applications, we encourage the community to develop *self-contained* applications that mitigate important security threats *by design*. We hope our work can shed light on the development of robust switch-based security applications and inspire developers to perform more rigorous security analysis.

## 2 BACKGROUND

**Terminology.** We use *data-plane* applications to refer to both software and hardware based data-plane applications. We use *switch-based* applications for the data-plane applications that are implemented on high-speed programmable *hardware* switches.

**Threat model.** A protected network deploys data-plane applications to process incoming and/or outgoing network traffic. We respect the original threat models (if any) for our target applications, and assume the adversary is capable of spoofing IP addresses for applications running on the network border. Additionally, we assume the adversary may also control a few hosts in the protected network. We focus on two types of attacks. In *disruption* attacks, the adversary's goal is to disrupt the the legitimate flows from/to legitimate clients. In other words, the adversary tries to increase the false-positive (e.g., marking legitimate flows as malicious) rate of the application to increase collateral damage and operational costs. In *evasion* attacks, the adversary tries to evade the detection of the application to make its false-negative (i.e., failing to detect malicious flows) rate higher.

**Related work.** Prior work has developed various program analysis/debugging/verification tools for data-plane applications (mostly P4-based) [5, 6, 16, 21, 21, 25, 36], and have reported implementation bugs and security vulnerabilities in many switch-based applications. Specifically, Kang *et al.* found that a set of switch-based applications are vulnerable to various performance degradation attacks [16].

Meier *et al.* [24] examined the security of data-driven systems against adversarial traffic, which covers several switch-based applications (and one concrete case study for Blink [13]). In contrast, our work focuses more on switch-based security applications, and our attacks exploit diverse attack vectors introduced by the switch data plane.

## 3 DESIGN CHALLENGES IN DATA-PLANE SECURITY APPLICATIONS

In this section, we first discuss fundamental limitations for in-network monitoring, whether implemented on software middleboxes or data-plane hardware. We then discuss the common problematic design patterns in switch-based applications. *We clarify that some of the design issues may not be specific to switch-based applications.* However, the restricted programming model and resources in the programmable switches increases the likelihood of developers adopting such problematic design patterns.

### 3.1 Limitations of in-network monitoring

Traffic monitoring is an essential component in data-plane (security) applications. One important factor that affects monitoring accuracy is where the monitor is located. Routing changes can prevent the monitor from seeing all of the packets of a flow, and asymmetric routing can limit the monitor to seeing just one direction of the traffic. However, even if the monitor can see all of the packets in both directions, other challenges remain:

Ⓐ **No effective detection of spoofed traffic.** In-network applications on the Internet are vulnerable to spoofing attacks. Most of the data-plane applications simply assume that all packets are legitimate, and packets of the same five-tuple belong to the same flow, ignoring the consideration that some (or all!) of the packets may be injected by an adversary (e.g., [20, 22, 32, 33, 35]). Different applications may have different levels of "tolerance" to spoofing. For applications with a limited attack surface (e.g., monitoring outgoing flows in an enterprise network where spoofing defenses are deployed) [38], spoofing may not be a concern. But applications that are open to adversarial (Internet) traffic may need to consider the performance and security implications of spoofing: the adversary can leverage spoofing to fill a flow table with useless entries (resource exhaustion), invoke slow-path processing continuously (performance degradation), manipulate flow statistics to make legitimate flows look malicious (disruption), or make malicious flows look normal to evade detection (evasion). Note that resource exhaustion and performance degradation are well studied in the prior work [16, 24], so we only focus on disruption and evasion attacks in our work.

Ⓑ **Insufficient understanding of network protocols.** Modern network protocols have tremendous complexity. Data-plane applications often consider simplified models for selected network protocols, while in practice heterogeneous end-hosts could use a diverse set of protocols, and the same protocol can have different extensions and variants (like TCP). One consequence of the mismatch between full protocol complexity and simplified model is a failure in modeling all the possible normal protocol behaviors. Therefore, the application may be too permissive and fail to identify malicious flows (evasion), or be too restrictive and identify many legitimate flows as malicious (high false positives). A prior study has shown that an adversary, without spoofing, may exploit the lack of knowledge of protocol behavior in NIDS to bypass NIDS [18]. Detecting and preventing such malicious behaviors (or unintentional errors) requires implementing *all* the necessary checks and state machines in the data plane if possible (or the control plane at the cost of performance), which are expensive for data-plane applications.

Ⓒ **No visibility into end-host state.** Data-plane applications generally perform analysis only on the network traffic (without considering end-host state). As a result, they do not know whether an observed packet will be accepted by the end-host successfully. For instance, malformed TCP packets (e.g., wrong sequence numbers) may be dropped by the receiving end-host, but may be considered to be a part of the connection by the data-plane application. One may argue that the lack of visibility of end-host state will not cause any security issues once challenge Ⓑ has been addressed. But, unfortunately, this is not always true. For instance, if the adversary flips bits in the encrypted TLS payloads and updates TCP checksums, the packets may look legitimate but will cause decryption errors at the end-hosts. The data-plane application cannot detect this without knowing end-host state. We will demonstrate how to exploit Ⓒ to evade security applications in §4.

Next, we will discuss how the most critical challenge—limited resources—in *switch-based* applications causes or exacerbates security problems.

### 3.2 Design pitfalls due to resource constraints

High-speed programmable switches have limited memory and processing resources, and a restricted programming model. This forces the designers of switch-based applications to explore a design space that is different from conventional software systems in order to fully leverage the performance benefits of switch data planes. The common compromises the designers make are:

- **Reduction**. The application may implement only the core components and ignore seemingly "nonessential" functionality, or fail to maintain some critical state, or omit certain operations for improving performance and reducing implementation efforts.

- **Approximation**: The application may replace a component (e.g., algorithm and data structure) with a data-plane-friendly component with similar properties. A common case is replacing a secure hash function with the built-in CRC hash functions of programmable switches.

With reduction and approximation, it might be easy to build a switch-based application with similar functionality as its software counterpart but it is very challenging for the application to be as robust as the software counterpart. Next, we discuss several pitfalls in switch-based application design caused by reduction and approximation and how an adversary may exploit them.

① **Use of coarse-grained key.** A common component in switch-based applications is *packet grouping*. In packet grouping, the application clusters packets of the same *packet key* (selected header fields) into packet groups, and perform operations (gathering statistics, rewriting IP addresses, etc.) in the same way for all packets in the same group. While the packet key is usually the five-tuple, some applications may prefer a more coarse-grained key, such as using only source IP address, to avoid maintaining per-flow state. As we will demonstrate in §4.1, the use of a coarse-grained key may facilitate evasion or disruption attacks.

② **Lack of packet/protocol semantic checking.** Applications tend to assume that any observed packets are legitimate and do not validate packet format or check whether the interaction between two end-hosts follows protocol specifications. Some simple checks could be useful for detecting adversarial traffic, e.g., one may want to further inspect the TCP packets or flows if they contain wrong checksums or out-of-window sequence numbers. Unfortunately, performing even simple checks could be expensive in programmable switches, or impossible because of limited support for parsing variable-length application-layer data. The lack of semantic checking not only introduces security exploits as discussed in Ⓑ, but may make spoofing attacks easier. The adversary needs to only ensure that the header fields checked by the application—usually just source/destination IP addresses and port numbers—look legitimate, without worrying about whether the other header fields are carrying the correct values.

③ **Discrepancy between implementation and specification.** When implementing an existing network or security protocol in programmable switches, the developer needs to carefully optimize the implementation so that it can fit the switch data plane: approximating certain functions with switch-friendly counterparts, simplifying some operations to cope with switch resource constraints, and omitting operations that are not feasible (e.g., loop). After reduction and approximation, the resulting implementation may deviate from the corresponding standards/specifications of the network or security protocol, and fail to handle various corner cases that could be security-critical. This could make the switch-based application more fragile in the presence of vagaries of network protocols and adversarial traffic.

④ **Use of insecure hash.** Cryptographic secure hash functions are essential building blocks for many security and privacy applications. Unfortunately, current programmable switches do not offer secure hash functions, and only provide CRC16 or CRC32 hashes instead. The short hash length is subject to the birthday bound [8] and is prone to collision. Moreover, unlike a secure hash function, the CRC family does not provide *preimage resistance*, i.e., hard to recover the input message based on the hash output. For a CRC32 function with known polynomials and one function output, the adversary only needs to try at most $2^{32}$ messages to find the corresponding input message. For a CRC32 function with unknown polynomials, the adversary can try at most $2^{32}$ polynomials to find polynomials being used, given one input message and the resulting function output. Exploiting certain mathematical properties of CRC (e.g., $CRC(A) \oplus CRC(B) = CRC(A \oplus B)$), there may be more efficient ways to break CRC [4, 7].

⑤ **Use of insecure data structures.** To prevent resource-exhaustion attacks, switch-based applications may use compact data structures with constant resource requirements. This gives a false sense of security. A recent study of the security of probabilistic data structures shows that bloom filters, counting bloom filters, and count-min sketches are all vulnerable to target-set coverage attacks, a new type of pollution attack, even if the underlying hash functions are independent perfect random oracles [3]. HyperLogLog and Cuckoo Hash are also vulnerable to pollution attacks and denial-of-service attacks (e.g., forcing insertion failures for certain elements) [28, 29]. The secure versions of these data structures require the use of salts, keyed hash functions, or large data structures (e.g., $\geq 2^{15}$ filter bits for bloom filters to mitigate the attacks). All of these secure data structures are hard to implement on programmable switches, without compromising performance. Most importantly, the essential building block, a secure hash function, is not available in the switch data planes.

The aforementioned attacks exploit hash collision. The data structures that approximate hash functions with CRC could be more vulnerable to these attacks because of the high collision probability, allowing the adversary to manipulate false positive or false negative rates with greater ease.

⑥ **Lack of protection for slow/critical paths.** Hardware/software co-design is common in switch-based applications. Certain complex and expensive tasks cannot be done in the switch data plane, and offloading the corresponding computations to the control plane or CPU becomes the only option, though the control plane is much slower than the switch data plane. Generally, co-designed applications may be vulnerable to slow-path attacks in which the adversary sends packets to invoke the control plane to disrupt application performance. This issue is well-known, but few applications have considered or implemented countermeasures, even as simple as rate limiting, to protect the slow path. Though several mechanisms have been proposed [16, 38], there is no thorough evaluation in the literature that demonstrates the effectiveness and efficiency of these mechanisms.

**Discussion.** In some cases, developers may even be aware of design pitfalls we enumerate. For example, developers are aware of the potential security issues introduced by approximating a hash function via CRC, and usually suggest replacing it with a stronger cryptographic secure hash function while still using CRC in their applications. However, implementing a secure hash function without significant performance degradation in the switch data plane is not trivial. We are not aware of any switch-ASIC-based implementation of secure hash functions.

| App | Insight | Attacks | Major Issues |
|---|---|---|---|
| NetHCF | Insiders can facilitate spoofing | D/E | ⒶⒸ① |
| NetWarden | Malformed packets can help shaping traffic | E | Ⓒ② |
| Syn proxies | Forge CRC-based cookies is easy | E | ⒷⒸ①③④ |
| P4RTT | On-path adversary can bypass detection via spoofing | D/E | ⒶⒸ② |
| P4Knocking | Coarse-grained key can be exploit to cause false positives | D | Ⓐ①② |

Table 1: A summary of examined switch-based security applications and the attacks they are vulnerable to. "D" and "E" stand for disruption and evasion respectively.

## 4 CASE STUDIES

Next, we demonstrate the broad applicability of the design challenges mentioned above through case studies of six switch-based security applications. Each case study showcases how adversaries can exploit these design challenges to compromise system security or disrupt system operations by introducing collateral damage.[1] In the worst cases, *our attacks can completely bypass the very protection that these switch-based applications were designed to provide*. We summarize our insights in Table 1. Note that for each application, we only highlight the major issues that we have exploited to develop our attacks; the other issues do exist in the applications, such as ⑤ and ⑥ in NetWarden and ⑥ in NetHCF, and the attacks proposed in prior work can still be applied [3, 16].

### 4.1 NetHCF (Issues: ⒶⒸ①)

NetHCF is a switch-based implementation of Hop-Count Filtering (HCF) [14], a mechanism for filtering spoofed IP packets [19]. HCF was developed based on the observation that an adversary may find it challenging to accurately spoof the number of hops a packet takes to reach its destination (i.e., hop counts). With HCF enabled, a server infers the hop-count information for a given IP address/prefix based on the Time-to-Live values in the IP headers, and maintains an IP to hop-count (IP2HC) mapping table to detect and block spoofed packets. When building/updating the IP2HC table, NetHCF considers the presence of adversarial traffic, and only collects hop-counts from "legitimate" TCP flows. NetHCF assumes that a flow is legitimate if it *sees* a correct three-way handshake (correct packet sequence and corresponding sequence and acknowledgment numbers) at the beginning of the flow.

**Attacks.** As discussed in Ⓒ, seeing the handshake packets does not mean the handshake is actually happening between end-hosts. Assuming that the adversary controls a host in the protected network, it is trivial to fake the three-way handshake with arbitrary source IP addresses. As illustrated in Figure 1, even if the adversary cannot receive the SYN-ACK from the controlled host $B$ in a handshake, the adversary can still send back a legitimate-looking ACK as if it actually received the SYN-ACK. Since the adversary controls $B$, the sequence number used in the SYN-ACK can be selected and set by the adversary. As in Figure 1 (B), NetHCF will assume the adversary flow ($A \leftrightarrow B$) is legitimate and use it to collect the hop-count for IP $A$. The adversary is able to insert a false hop count for $A$ into the IP2HC table. Following the attack, packets from the real host at $A$ will look incorrect to NetHCF and get blocked (i.e., disruption attacks), and packets from spoofed IPs will look legitimate (i.e., evasion attacks).

---
[1]Our attack strategies may also work for other switch-based or non-switch-based applications of similar designs as the vulnerable applications.

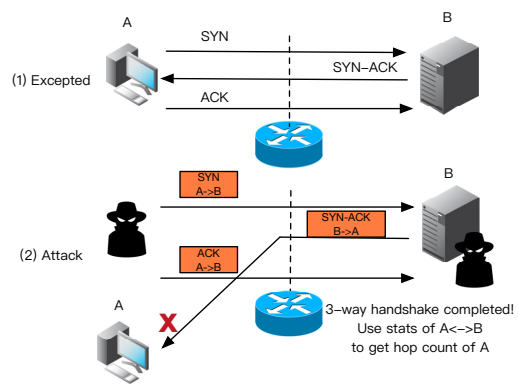

Figure 1: The adversary uses a host in the protected network to "simulate" TCP handshakes between the controlled host and spoofed IP addresses to bypass NetHCF TCP monitoring. The orange color boxes represent spoofed packets.

End-host-based HCF constructs the IP2HC mapping table implicitly per source-destination IP address pair. Even if the adversary manages to pollute the IP2HC table entry for an IP $A$ at host $B$, only the flow $A \leftrightarrow B$ would be affected. However, NetHCF constructs IP2HC table on a per source IP address basis to save memory. This reduction (①) exacerbates the damage caused by disruption attacks because all flows from $A$ to the network are blocked.

### 4.2 NetWarden (Issues: Ⓒ②)

NetWarden performs covert timing and storage channel (i.e., exfiltrating data via packet timing or header fields) detection using programmable switches [38]. It compares the inter-packet delay (IPD) distributions of monitored flows with expected distributions to detect timing anomalies, and validates and rewrites selected IP/TCP headers (e.g., IPID) to mitigate potential storage channels. NetWarden is deployed on a Top-of-Rack (ToR) switch in data center, where IP spoofing is restricted, and monitors the outgoing flows from the (compromised) servers.

**Attacks.** NetWarden does not validate packet fields that related to packet payload such as checksum and payload length (②). The support for accessing variable-length payload is limited in programmable switches [37]. Therefore, timing-based covert channels can leverage malformed packets (e.g., incorrect TCP checksums) to bypass NetWarden. The adversary can also use malformed packets to shape IPD distributions. Assuming that two packets with an IPD of one second are used to encode "1", by sending one malformed packet in between the IPD observed by NetWarden becomes 500 ms.

The receiving end simply ignores the malformed packets (may already be dropped by middleboxes) when computing IPD. However, the receiving end can still send back responses as if it has received the correct packets (Ⓒ).

## 4.3 SYN proxy (Issues: ⒷⒸ①③④)

SYN proxy is a defense against TCP SYN flood attacks. The SYN proxy responds to a SYN with a SYN-ACK, of which the initial TCP sequence number is a challenge (i.e., SYN cookies) generated by a secure hash function. If the ACK (for SYN-ACK) sent from the client contains the correct challenge, the proxy allowlists the corresponding five-tuple and forwards packets back and forth between the client and server. Several switch-based SYN proxies have been proposed and implemented [23, 30, 40].

**Attacks.** One SYN proxy design [30] uses CRC16 to generate SYN cookies: $cookie = \text{CRC16}(\text{source IP}, \text{source port}, C)$ where $C$ is a constant value (③). Note that the flows with the same client source IP address and port number get the same cookie value, while the original SYN proxy generates per-flow cookies (Ⓑ). To get $C$, the adversary just needs to send a SYN and extract the cookie from the returned SYN-ACK. It is easy to brute force $C$ given the cookie and source IP address and port (④). The design does not track TCP connections; it only checks the ACK of the SYN-ACK in a handshake to avoid maintaining per-flow state, and allowlists the source and destination address pair (instead of five-tuple) once the ACK has the correct cookie value (①). Thus, to evade the defense, the adversary just needs to send a number of spoofed ACKs with the correct cookie values (as if the attacker actually received SYN-ACKs) (Ⓒ), which are computed based on via CRC16. Once a spoofed IP address get allowlisted, the adversary can launch SYN flooding (or other) attacks with the spoofed IP address.

The SYN proxy provided by Jaqen adopts a slight different design, in which the cookie returned from the client will be in a RST packet (please see the original paper for more details) [23]. However, similar to the first SYN proxy, Jaqen's SYN proxy only check RSTs to avoid maintaining per-flow state, and uses CRC32 for cookie generation: $cookie = \text{CRC32}(\text{five-tuple}, \mathcal{N})$ where $\mathcal{N}$ is a nonce reused for a given time epoch across all flows (③, ④). The adversary can easily recover the nonce, craft the correct cookie for any given five-tuple, and send RSTs to evade the defense.

We acknowledge that the two SYN proxies discussed above do increase an attacker effort. Let $N$ be the number of SYN packets the adversary needs to send for a successful SYN flooding attack against a victim server. For the first proxy, the adversary needs $N + N/2^{16}$ packets—using $N/2^{16}$ ACKs to get $N/2^{16}$ IP address allowlisted, and varying source port number to send $2^{16}$ SYNs for each IP address. For the second proxy, the adversary needs $2N$ packets. For each SYN of a given five-tuple, the adversary sends a RST before the SYN to get the five-tuple allowlisted. Nevertheless, the extra effort could be still acceptable for a dedicated adversary.

We suspect that the SYN proxy proposed in Poseidon has similar security issues as the other two proxies because it also uses CRC for hashing [40]. Unfortunately, we do not have the source code and sufficient details from the paper for us to evaluate its design. Nevertheless, the Jaqen paper [23] points out that Poseidon is vulnerable to hash collision (④), which can be exploited to launch disruption attacks. We refer the readers to Jaqen [23] for more details.

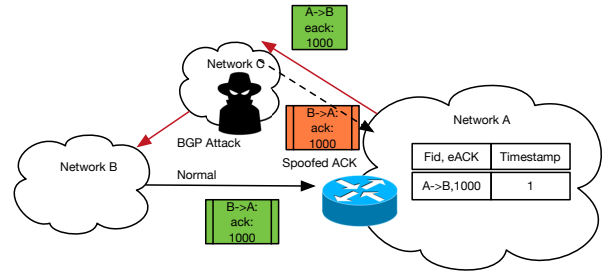

**Figure 2: The adversary strategically sends spoofed ACKs to P4RTT to pollute RTT measurement results to bypass BGP hijacking detection.**

## 4.4 RTT monitoring (Issues: ⒶⒸ②)

P4RTT runs on a border switch and measures round-trip times (RTTs) in the data plane [2]. R4RTT monitors each outgoing TCP data packet, and matches the associated return ACK based on the acknowledgment number. Once it finds a match, P4RTT removes the corresponding record from its table and calculates the RTT. One security use case of P4RTT is to detect BGP attacks: since BGP attacks typically change the routing paths of normal traffic, the attack may introduce higher round-trip times.

**Attacks.** We argue that a dedicated adversary who is capable of launching BGP attacks can bypass P4RTT detection with low additional costs. Let us suppose that the adversary is able to intercept the traffic from the protected networks $A$ to a network $B$, as in Figure 2. For a TCP packet from $A$ to $B$, the adversary can craft the corresponding ACK packet with the correct ACK number, and send it to $A$ from a strategically selected network $C$ before the legitimate ACK reaches $A$ (or delay the TCP packet). $A - C$ and $A - B$ should have similar RTTs. P4RTT will use the first arriving ACK to compute the RTT, and remove the matching record. The legitimate ACK received later will be ignored. The RTTs of a flow under BGP attacks will still look normal to P4RTT. Even though the end-hosts may receive malformed ACKs and might be aware that they are under attack, P4RTT will not be aware of the maliciously injected packets and take appropriate action, because of the lack of visibility into end-host state and protocol semantic checking (Ⓒ②). Similarly, the adversary can also delay the normal TCP packets to produce inflated RTTs, launching disruption attacks.

## 4.5 P4Knocking (Issues: Ⓐ①②)

P4Knocking is a switch-based port-knocking-based authentication system [39]. In port knocking, the client needs to send a sequence of SYNs to the right sequence of ports of the server before initiating the real connection. If any of the SYNs knocks a wrong port, the connection attempts from the client will get blocked. To reduce state, P4Knocking tracks the knocking sequence per source and destination IP address pair.

**Attacks.** To disrupt the connections from a victim IP address to a server in the protected network, the adversary can send spoofed SYNs with the victim IP address as the source address to random ports of the server. If some connections from the victim IP address are performing port knocking, the spoofed SYNs will be injected into their knocking sequences as observed by P4Knocking, resulting in incorrect knocking sequences. Subsequently all the future connections from the victim IP address to the server will be blocked for a given time period.

## 4.6 P4DNS (Issues:Ⓑ②③)

Finally, we discuss a non-security application to demonstrate the difficulty of implementing a standard protocol correctly in programmable switches. P4DNS offloads DNS to programmable switches to achieve 50x performance improvement over software-based DNS servers [37]. To speed up DNS resolution, P4DNS stores the response for a domain name in a DNS cache table in the data plane.

**Attacks.** DNS transaction ID (TXID) is used to associate a DNS response with a DNS request to mitigate DNS response spoofing. Unfortunately, P4DNS assumes all DNS responses are legitimate and does not check TXID. The adversary can easily spoof a DNS response with an arbitrary TXID that associates a domain to malicious IP addresses, and pollute the DNS cache in P4DNS.

## 5 ONGOING WORK AND CONCLUSION

In this work, we enumerated the major design challenges and common design pitfalls in developing switch-based applications. We developed concrete attacks against six switch-based security applications to showcase how adversaries can exploit these design issues to completely bypass the protection these application were designed to provide or disrupt their normal operations. As future work, we will provide a more thorough analysis on side-channel vulnerabilities and false alarms in a broader set of switch-based applications, and explore how to mitigate these vulnerabilities.

### 5.1 False alarms for benign traffic

We define false alarms as *benign* flows being disrupted or dropped by an application in the absence of any attacks. High false alarms pose a challenge to deployability. Even a low rate of false alarms may impact a significant number of users, considering that the application may process a high volume of traffic. Insufficient understanding of network protocols (Ⓑ) is a common cause of false alarms. We discuss false alarms in NetWarden in this work, and will continue examining other switch-based security applications in future work.

**NetWarden.** NetWarden highlights a feature called *ACK boosting* for performance improvement. When ACK boosting is enabled, NetWarden crafts and sends TCP ACKs to the server on behalf of the client, so the server thinks the client has already received the data, and will immediately send its data. The real ACKs sent from the client will be dropped by NetWarden. We collected 30-min campus TCP traffic traces ( 26 GB), and checked the ACKs in the HTTP and HTTPS flows (0.65 M) in our traces. We found 17.9% of the ACKs (6.1 M out 34.0 M) contain data, corresponding to 18.1% of the HTTP(S) flows. Thus using NetWarden with ACK boosting may disrupt 18% of the HTTP(S) flows.

### 5.2 Side-channels in data-plane applications

Data-plane applications usually serve the entire network and aggregate information across all flows. This may give the adversary an opportunity to learn information about the network or individual hosts in the network, which can have privacy implications. Considering that data-plane applications may become the easy attack targets—running at the network border but without sufficient protection—reducing the attack surface for side-channel attacks in the design phase can be critical for security and privacy.

Multiple levels of timing side-channels exist in data-plane applications. Differences in packet parsing and processing piplines, packet recirculation, control plane invocation, and action on packet or flow (e.g., dropping and rerouting) can all introduce measurable delays, serving as signals for side-channel attacks. Depending on the applications, the adversary can learn different information. For example, in NetCache (an in-network key-value store) [15], the adversary can infer the hottest items on a server by sending queries to different items and checking the response time. A query will be handled by the switch if the queried item is cached; otherwise, it will be processed by a server, resulting in a longer response time. Another example is that of a hypothetical application which detects heavy hitters and reroutes these flows. Let us consider an adversary who knows that a user is likely to visit a web server (with a public IP address) in the protected network. The adversary can spoof packets using the user's source IP address (and brute force port numbers), and infer when the rerouting happens based on some latency measurements (e.g. [11]). Based on the number of spoofed packets required for triggering rerouting, the adversary may infer whether at a given time the user is visiting the server, or estimate the number of packets/bytes the user has sent. We plan to systemically investigate side-channels in data-plane applications.

### 5.3 Preliminary strategies toward developing a secure switch-based application

Beyond using program analysis, debugging, and verification techniques [5, 16, 21, 25, 36], a complementary approach is *black box testing* (e.g., fuzzy testing). Black box testing does not require access to the source code. An automated black box testing framework is useful for understanding the patterns of adversarial traffic that applications are vulnerable to and discovering vulnerabilities caused by logical loopholes. Recall that performance optimizations in data-plane applications may result in discrepancy between implementation and specification (③). Therefore, another strategy is to leverage existing robust and secure software counterparts that can be used as *reference implementations* for some data-plane applications. Similar to data-plane equivalence checking [6], one may perform data-plane and general software application equivalence checking. For instance, translating a data-plane program to other language like C (or in the reverse direction) [27], and compare the translated program against the corresponding reference implementation to examine resulting discrepancies.

Finally, we note that current open-source P4 applications are usually monolithic and tightly-coupled, which makes reusing a component, such as a data structure and algorithms, developed in prior work more difficult. Using higher-level languages than P4 [9, 10, 12, 31, 34] could facilitate the development of modular code. We encourage the community to focus on developing reusable, open-source reference implementations of common components, and to scrutinize them together.

Overall, we hope our work can motivate the community to prioritize security and develop switch-based applications with rigorous security by design.